# Open Source Codes for Computing the Critical Current of Superconducting Devices

Víctor M. R. Zermeño, Salman Quaiyum, Francesco Grilli

*Abstract*—In order to transport sufficiently high current, high-temperature superconductor (HTS) tapes are assembled in cable structures of different forms. In such cables, the tapes are tightly packed and have a strong electromagnetic interaction. In particular, the generated self-field is quite substantial and can give an important contribution in reducing the maximum current the cable can effectively carry. In order to be able to predict the critical current of said cable structures, a static numerical model has been recently proposed. In this contribution, we present in detail the implementation of such models in different programming environments, including finite-element-based and general numerical analysis programs, both commercial an open-source.

A comparison of the accuracy and calculation speed of the different implementations of the model is carried out for the case of a Roebel cable. The model is also used to evaluate the importance of choosing a very accurate description of the angular Jc(B) dependence of the superconductor as input for the material's property.

The numerical codes, which are open-source, are made freely available to interested users.

## I. INTRODUCTION

DEVICES BASED on high-temperature superconductor (HTS) tapes are often designed with a tight arrangement of the tapes, in order to maximize the engineering current density. Cables designs such as the Roebel cable [1], the conductor-on-round-core (CORC) cable [2], and the twisted stacked-tape cable (TSTC) [3] have a different geometrical layout, but they are all characterized by a compact arrangement of the strands. This leads to a strong electromagnetic interaction between the strands, and the determination of the self-field critical current of the cables is not trivial, particularly taking into account the fact that the critical current density $J_c$ of HTS coated conductors (CC) often exhibits a complicated angular dependence with respect to the local magnetic flux density $\boldsymbol{B}$ [4][5].

Recently, we have proposed a new method to quickly calculate the critical current of HTS cables and windings, using the $J_c(\boldsymbol{B})$ dependence of the superconductor as input [6]. The method uses a power-law $E$-$J$ relationship for the superconductor. The strength of the method resides in the introduction of a variable $P$ that allows avoiding the direct solution of the nonlinear $E$-$J$ relationship, hence providing a simpler problem that can be easily solved. Another remarkable feature is that, depending on the application considered (cable, coils, coils made of cables) and the particular experimental setup, different criteria for the definition of the critical current $I_c$ can be used. The method has been successfully tested against experiments for a variety of superconducting devices and working conditions.

With this contribution we aim to explain in more technical details how the model works in a variety of software implementations, including finite-element programs, such as FreeFem++ [7][8], and more general programs for numerical analysis, such as Matlab [9] and GNU Octave [10]. For completeness, the results obtained are compared with a previous implementation reported in [6] made using Comsol Multiphysics [11]. In the appendix of this article, the corresponding open source codes – namely FreeFem++ and GNU Octave – are made freely available to the readers. Given its verbosity, the scripted COMSOL model was considered to be too long to be presented and discussed in a journal paper. Nevertheless, all codes (COMSOL included) are available for download at the MODEL FILES section of the HTS modelling workgroup website [12].

The reason behind our choice is to give a tangible contribution to bypass one of the obstacles toward a more rapid advancement of the field of numerical modeling of superconductors: the availability of common codes and the possibility of sharing them [13].

The paper is organized as follows: section 2 contains a brief description of the model and its different software implementations. Section 3 contains the results for the case of a Roebel cable composed of ten HTS coated conductor strands characterized by anisotropic $J_c(\boldsymbol{B})$ dependences: first, the model implementations are compared in terms of results for the critical current calculation and speed; then, the model is used to evaluate the necessity of choosing a very precise $J_c(\boldsymbol{B})$ characteristic to reproduce the often complicated angular dependences introduced by artificial pinning processes. The conclusion summarizes the main contributions of this work, while the appendix contains the publicly available codes of the various software implementations.

Automatically generated dates of receipt and acceptance will be placed here; authors do not produce these dates.

This work has been partly supported by the Helmholtz Association (Young Investigator Group grant VH-NG-617).

V. M. R. Zermeño (e-mail: victor.zermeno@kit.edu), F. Grilli (e-mail: francesco.grilli@kit.edu), and (until March 2015) S. Quaiyum (email: salman.quaiyum@gmail.com) are with the Institute for Technical Physics, Karlsruhe Institute of Technology, Hermann-von-Helmholtz-Platz 1, 76344 Eggenstein-Leopoldshafen, Germany.

## II. Description of the models

A complete description of the model in its differential form is given in [6]. For completeness, it is briefly described here. The model solves Ampere's law with current sources whose amplitude depends on the local magnetic flux density $\mathbf{B}$. For a given superconducting strand, this dependence is assumed to be proportional to the critical current density $J_c(\mathbf{B})$. Considering all currents being perpendicular to the $x$-$y$ plane, the governing equation can be written in terms of the $z$ component of magnetic vector potential $A$ as:

$$\nabla \cdot \nabla A_z + \mu_0 J_c(\mathbf{B}) P = 0, \qquad (1)$$

where $\mathbf{B} = \nabla \times \mathbf{A}$. $P$ is a domain-wise uniform variable and it is used to enforce a given current amplitude in each strand. In the case of transposed cables, this amplitude is the same in each strand. As discussed in [6], when the relation between the electric field $E$ and the current density $J$ is described by a power law with exponent $n$, $E$ is given by $E = E_c P / P^{n-1}$. Here $E_c$ is the electric field at when the current density has reached its critical value $J_c$.

Using an alternate modeling formulation, each conductor in the cable can be assumed to be a discrete collection of lines of current. If said discretization is dense enough, the local magnetic flux density at the location $(x_i, y_i)$ of the $i$-th line of current can be approximated as the one produced by all the other neighboring lines of current in the cable. Using Biot-Savart, it can be expressed as:

$$B_i = \frac{\mu_0}{2\pi} \sum_{j \neq i} P I_c(B_j) \frac{\{-(y_i - y_j), x_i - x_j\}}{(x_i - x_j)^2 + (y_i - y_j)^2} \qquad (2)$$

Here, the net transport current in the $j$-th line of current $I_{t,j}$ is proportional to the local critical current $I_c(\mathbf{B}_j)$ so that $I_{t,j} = P I_c(\mathbf{B}_j)$. For a strand of width $sw$ and thickness $th$, which has been discretized with $m$ lines of current, the local critical current $I_c(\mathbf{B})$ is defined as $I_c(\mathbf{B}) = sw \, th \, J_c(\mathbf{B}) / m$.

Either formulation (1) or (2) is solved iteratively provided a criterion to determine the critical current of the cable. In the particular case of transposed cables, such as Roebel, its critical current can be defined using two criteria [6]:

1. The current at which at least one conductor in the cable reaches a voltage drop per unit length equal to the critical value $E_c$. This is referred to as the MAX criterion.
2. The current at which the average electric field has reached its critical value $E_c$. This is referred to as the AVG criterion.

The interested reader is directed to [6] for an additional understanding of the model's capabilities and applicability for estimating the critical current of single tape coils, coils made of transposed cables, non-transposed cables and cables with ohmic termination resistances.

Although both formulations (1) or (2) are in principle equivalent, there are significant differences regarding their usability. For example, when solved using the Finite Element Method (FEM), (1) can easily take into account arbitrary geometries with ease as all the geometry creation and meshing is handled automatically regardless of the shape of the conductor. In the same manner, if written in its more general form, $\nabla \cdot (\mu^{-1} \nabla A) = J_c(\mathbf{B}) P$, it can also include the effect of magnetic materials. This allows for example calculating the critical current of multi filamentary wires embedded in a magnetic material matrix [14].

On the other hand, although (2) cannot incorporate magnetic materials, it can be solved very rapidly as it only considers the superconducting regions of the cables. Having a very large aspect ratio, Coated Conductor (CC) tapes can be approximated as a collection of coplanar lines of current. Therefore complex cables composed of CC tapes such as the Roebel can be easily and accurately represented and their critical current calculate rapidly calculated using (2).

In an effort to present the code and discuss it within this work, the models are written in a very compact form. Although the code is fully commented, the following subsections give a general description of both implementations (1) in FreeFEM++ and (2) in MATLAB/Octave.

Both models are very versatile and can be easily tailored to simulate different devices or conditions of operation. For instance, externally applied field can easily be included by appropriately modifying the boundary condition `Az=0` in line 30 of the FreeFEM++ code. In a similar manner, an externally applied field $\{Bx_{ext}, By_{ext}\}$ can also be implemented in the MATLAB/Octave code by changing `Bx` to (`Bx+Bx_ext`) and `By` to (`By+By_ext`) in line 25.

**FIG. 1 HERE**

**TABLE I HERE**

### A. Free FEM ++

FreeFem++ is a partial differential equation solver that is freely distributed under a LGPL License. One interesting feature of FreeFem++ is that in addition to the multiplatform installations available for download [7], it is available for use on the web without the need to even set an account [15].

The FreeFem++ code is presented in section A of the appendix. It is composed of just 48 lines. The first 27 lines deal with the header, declaration of parameters, variables, functions, and with the creation of the geometry and the mesh. The following 17 lines deal with stating the problem in weak form (lines 29 and 30) and its solution using an iterative solver. The post processing is done in the last 3 lines of the code. The code is in fact compact, excluding comments and complementary post processing, 37 lines are enough to calculate the critical current of a Roebel cable.

From the user point of view, the necessary parameters need to be input in lines 4 to 7. The critical current criterion is set with the variable `s` (line 4). The parameters related to the particular $J_c(\mathbf{B})$ expression used are input in line 5. Other expressions can be easily implemented in line 26 to account for superconductors with different $J_c(\mathbf{B})$ characteristics. The remaining input parameters (lines 6 and 7) are given in Table I and shown in Fig 1.

The problem solution is carried out in two nested stages. First, an initial estimate `I0` for the critical current of every strand of the cable is used to find a self-consistent solution for the problem described in lines 29-30. This is done iteratively in lines 34-41. The process involves estimating the magnetic field produced by a given current density. This magnetic field is then used to evaluate $J_c(B)$ locally in the tapes and with this obtain a new current density. The process is repeated until the error (defined by the $L^\infty$-norm) between the previous (`Az0`) and the new (`Az`) estimates for the magnetic vector potential is below a given tolerance (`tolAz`). The second stage iteratively updates the value of the transport current in each strand of the cable until the desired criteria for the critical current is met – within a given tolerance value (`tolp`).

In the post processing section, the critical current of the cable is calculated. For completeness, a plot of the magnetic flux density is given as shown in Fig 1.

### B. Matlab/Octave

Matlab is available for purchase at the mathworks website [9] under several licensing options. GNU Octave is distributed under the terms of the GNU General Public License and is available for free download at [10].

The Matlab/Octave code is presented in the appendix B. It is composed of just 44 lines. The first 10 lines include the header, and the declaration of parameters and variables. Lines 11 to 15 are used to create an array of points were the lines of current are located. Lines 16 to 19 are used to create auxiliary variables defined as $r2=(x-x')^2+(y-y')^2$, $xn=if(r2>0,(x-x')/r2,0)$ and $yn=if(r2>0,(y-y')/r2,0)$. The vector *{-yn, xn}* corresponds to the right factor on the right hand side of equation (2) and is later used in lines 28 and 29 to calculate the magnetic field.

The iterative solver (lines 20-34) is composed of two nested while loops. The first one checks whether the $I_c$ criteria selected is met (line 21) within a given tolerance "`tolp`". The second loop (lines 23-31) ensures that the calculated values for magnetic field and critical current are self-consistet within a given tolerance "`tolIc`". Lines 25 to 29 are in essence the implementation of equation (2) so that it can be solved iteratively. The value of the electric field "`E`" and the net current in the strand "`I0`" are updated at the end of the $I_c$ criteria loop (lines 32-33).

Post Processing takes up lines 35 to 44. Besides providing the calculated critical current value for a given $I_c$ criterion, estimates for the average electric field and maximum $P$ value are provided. Finally, a line plot showing the calculated current densities in the superconducting strands of the cable is given.

From the user point of view, the necessary parameters need to be input in lines 5 to 7. As with the FreeFem++ code, the critical current criterion is set with the variable `s` (line 5). The parameters related to the particular $J_c(B)$ expression used are input in line 6. Other expressions can be easily implemented in line 25 to account for superconductors with different $J_c(B)$ characteristics. The remaining input parameters (line 7) are given in Table I and shown in Fig 1.

**TABLE II HERE**

### III. EXEMPLARY RESULT

#### A. Comparison of the different models

In order to compare the models, we calculated the self-field critical current of a 10-strand Roebel cable with the same geometrical and physical parameters as that described in section III.A of [5]: the cable is composed of 1.8 mm-wide strands and the superconductor has an elliptical $J_c(B)$ dependence. For each model, the critical current was computed both with the MAX and AVG criteria. As displayed in table II, the critical currents calculated with the different models practically coincide, the maximum difference being 0.4% of the average value. All computations were performed using a standard workstation computer (Intel i7 4960K, 6 cores at 3,6GHz, RAM 64GB). The MATLAB/Octave code is extremely efficient running in fractions of a second (~0.1 s). The Comsol implementation requires 6 seconds (once the model is loaded in the GUI, which can take up to 15 s). The FreeFEM++ code (which does not require loading a GUI) runs in a little more than half a minute (34 s - 45 s). Being the fastest implementation, the code in MATLAB/Octave is an ideal candidate for optimization purposes. On the other hand, even though the code in FreeFEM++ takes a little more to run, its compactness as a script and ease in the implementation, make a good case considering that it is freely available.

As expected, the MAX criterion gives a lower estimate value of $I_c$. In contrast to the case of coils described in [6], however, the difference between the two criteria is minimal, because in the considered geometry (straight Roebel cable) all the tapes are subjected to a rather similar electromagnetic environment.

**FIG. 2 HERE**

#### B. Influence of $J_c(B)$ accuracy on $I_c$ calculation

The models developed here use a known $J_c(B)$ expression for the superconducting material to calculate the effective critical current of a cable made of several tapes of such material. How to extract a sensible $J_c(B)$ expression from measurements? The experimental data usually consists in the characterization of the angular dependence of the critical current $I_c$ on the applied magnetic field. For applications where self-field produced by the current in an individual tape is a significant fraction of the total field, obtaining an expression for $J_c(B)$ cannot be achieved by a simple fit of the experimental data as the self-field effects will have a non-negligible contribution. In general, it is required to solve the inverse problem of finding an expression for $J_c(B)$ such that the experimentally measured values for $I_c$ can be reproduced when the self-field contributions are included. This topic has been and still is the subject of research by several groups [4][5][16][17]. Obtaining a precise description of the angular dependence of $I_c$ can be a non-trivial task, especially in samples characterized by artificial pinning, for which the angular dependence of $I_c$ exhibits peaks and valleys for angles different from the superconductor's crystallography's axes and lacks symmetry [4]

On the other hand, it is not clear yet how precisely one needs to reproduce the angular dependence of individual tapes in order to get a reasonably good estimation of the critical current of a cable made of those tapes.

Our model allows quickly testing different angular dependences and assessing the impact they have on a cable's critical current.

As an example, we considered a 10-strand Roebel cable assembled from 12 mm-wide tape exhibiting a complex angular dependence (Fig. 2). Being able to precisely reproduce this dependence require the use of a complex multi-parameter $J_c(B)$ expression. In the figure, the thin lines are the calculated critical currents of a single tape in the presence of background field with the procedure described in [18], which includes the self-field.

**FIG. 3 HERE**

The question we want to answer is: what is the error one makes by neglecting the angular dependence and considering – for example – only the dependence on the magnitude $|B|$ of the magnetic field (Fig. 3)? When the model for the single tape is run with the simplified $J_c(|B|)$ of Fig. 3, the constant values indicated by the thick horizontal lines in Fig. 2 are obtained. These average critical current values are often quite different from the measured ones (with over- and underestimations that depend on the angle, but can reach 52% in the sample here considered). Nevertheless, when one calculates the critical current for the corresponding Roebel cable, where the generated magnetic field has a variety of orientations, the over- and underestimation errors compensate and the critical current is within 4% of the value calculate with a more precise $J_c(B)$. The values calculated with the two dependences and with the MAX and AVG criteria are listed in Table III.

One other thing to keep in mind is that in this kind of calculation one assumes that the $J_c(B)$ dependence is the same for all the superconducting tape used in a cable. While reports on the longitudinal uniformity of the self-filed critical current $I_c$ of long piece of tape abound, the same cannot be said on the uniformity of the $J_c(B)$ dependence. If the $J_c(B)$ changes significantly along the tape, a precise extraction of these properties from a short sample might have little sense.

With the example analyzed and discussed above, we do not want to undermine the importance of the angular dependence. Rather, we want to point out that the extraction of a precise very precise $J_c(B)$ dependence from experimental data should also be accompanied with data on the uniformity of such dependence.

**TABLE III HERE**

*C. Estimation of AC losses*

It is well known in the literature that the self-field AC loss of Roebel cables lies between the estimates given by Norris for the strip and the ellipse [19][20][21][22]. Therefore, once the critical current is known, the AC losses can be estimated easily following said principle. Although this is not a general rule that could be simply extrapolated to other configurations, at least, in the case of Roebel cables it provides a ballpark figure that can be used in initial stages of cable designs.

IV. CONCLUSION

In this work we have presented 2 different open source codes to calculate the critical current of superconducting devices. For the case of Roebel cables considered here, the implementations are very fast. In particular, the Matlab/Octave code running in fractions of a second can be easily used to perform optimization. Although the FreeFem++ code is not as impressive in terms of computational speed, it is very flexible, allowing cables made with other superconducting conductors – not just CC tapes – to be considered.

The model was used to compare the effect that a precise angular dependent $J_c(B)$ and a simplified $J_c(|B|)$ have on the critical current of a 10-strand Roebel cable. It was found out that both expressions gave very similar estimates for the overall critical current of the cable.

In line with the idea of facilitate the advancement of the field of numerical modeling of superconductors, and hoping to be followed by others, all codes used are freely available for download.

APPENDIX

*A. FreeFEM++ code*

```
1  //    A FreeFem++ code to calculate the Ic of superconducting cables    //
2  // By Victor Zermeno and Salman Quaiyum   doi:10.1109/TASC.2015.XXXXXXX //
3  //Declaration of parameters and variables for geometry, Physics and mesh //
4  bool s=0; string c="AVG"; if(s){c="MAX";}; //Ic criteria: s=(1->MAX,0->AVG)
5  real Jc0=4.75e10, Bc=35e-3, b=0.6, k=0.25;         // Jc(B) parameters
6  int ns=10, ny=ns/2;                         // ns=number of strands in cable
7  real th=1e-6, sw=1.8e-3, rg=4e-4, sg=1e-4, n=21, tolAz=1e-9, tolp=1e-9;
8  real I0=Jc0*th*sw, x0=-sw-rg/2, y0=-((th+sg)*ny-sg)/2, E=0, Ec=1e-4, err;
9  real[int] XC(ns), YC(ns), Ics(ns), p(ns^2), pn(p.n); p=0.9; pn=0.9;
10 int[int] cm(1), hm(ns), vm(ns); cm=50; hm=-50; vm=-1;    // Mesh parameters
11 ///////////////////// Creation of geometry and mesh /////////////////////
```

```
12  for(int i=0; i<ns; i++){XC[i]=x0+(rg+sw)*(i/ny); YC[i]=y0+(sg+th)*(i%ny);}
13  border bb    (t=0, 2*pi) {x=20*sw*cos(t);    y=20*sw*sin(t);    label= 1;}
14  border top   (t=0, 1; i) {x=XC[i]+t*sw;      y=YC[i]+th;        label=i+2;}
15  border right (t=0, 1; i) {x=XC[i]+sw;        y=YC[i]+(1-t)*th;  label=i+2;}
16  border bottom(t=0, 1; i) {x=XC[i]+(1-t)*sw;  y=YC[i];           label=i+2;}
17  border left  (t=0, 1; i) {x=XC[i];           y=YC[i]+t*th;      label=i+2;}
18  mesh Th=buildmesh(bb(cm)+top(hm)+right(vm)+bottom(hm)+left(vm));
19  //////////////////////// Build FEM Solution Space /////////////////////////
20  fespace Vh(Th,P2);                          // Quadratic elements for Az
21  Vh Az, Az0, v;
22  fespace Wh(Th,P1dc);        // Piecewise-linear discontinuous elements for J
23  Wh J=Jc0;
24  p[Th(0,0).region]=0;                        // p=0 in the Air region
25  ////////// JcB and J as functions of Az using the dummy variable u /////////
26  macro JcB(u) Jc0/(1+sqrt((k*dy(u))^2+(-dx(u))^2)/Bc)^b //
27  macro J(u) JcB(u)*(p[region])//
28  //PDE(in weak form) Div(Grad(Az))+mu0*Jc(B)*p=0 and boundary condition Az=0
29  problem Pmodel(Az,v)=int2d(Th)(dx(Az)*dx(v)+dy(Az)*dy(v))
30                     -int2d(Th)(4e-7*pi*J(Az0)*v)+on(1,Az=0);
31  ////////////////// Solution using an iterative solver ///////////////////
32  while(abs(p.max-1)*s+abs(pn(0:ns:1).sum/ns-1)*(1-s)>tolp){  // Ic criterion
33      err=1;                                    // Reset err variable
34      while(err > tolAz){                       // Self consistency loop
35          Az0=Az;                               // Update old Az estimate
36          Pmodel;                               // Run FEM problem
37          for(int j=0; j < ns; j++){            // Ic and p value of j-th strand
38              Ics(j)=int2d(Th,j)(JcB(Az));
39              p(j)=I0/Ics(j);}
40          Az0=Az-Az0;         // Difference between old and new Az estimates
41          err=Az0[].linfty;}  // Error defined using the L-infinity norm
42      for(int i=0; i<ns; i++){pn(i)=p(i)^n;}
43      E= pn(0:ns:1).sum/ns*Ec;              // Average electric field in cable
44      I0=2*I0/((1+p.max)*s+(1+(E/Ec)^(1/n))*(1-s));}//Net current in strand
45  /////////////////Post Processing: Output data and plotting////////////////
46  cout <<endl<<endl<< "Ic= " << ns*I0 <<" A"<<" ("<< c <<" criteria)"<<endl;
47  Vh B=sqrt(dx(Az)^2+dy(Az)^2);
48  plot(B, bb=[[-1.1*x0,-1.1*y0],[1.1*x0,1.1*y0]], wait=1, fill=1, value=1);
```

B.  *MATLAB/Octave code*

```
1  %       A MATLAB code to calculate the Ic of superconducting cables     %
2  % By Victor Zermeno and Salman Quaiyum    doi:10.1109/TASC.2015.XXXXXXX %
3  %%      Initialization and declaration of parameters and variables     %%
4  clc; clear all; close all;
5  s=0; c='AVG'; if (s==1) c='MAX';end;     %Ic criteria: s=(1->MAX,0->AVG)
6  Jc0=4.75e10; Bc=35e-3; k=0.25; b=0.6;                  % Jc(B)parameters
7  m=100; ns=10; th=1e-6; sw=1.8e-3; rg=4e-4; sg=1e-4; n=21;    % parameters
8  mu0=4e-7*pi; Ec=1e-4; tolIc=1e-9; tolp=1e-9; %mu0, Ec criterion, tolerances
9  I0=Jc0*sw*th; P=0.5*ones(1,m*ns); E=0;    % Initial values for I0, P and E
10 [Bx,By,Ic]=deal(zeros(1,m*ns)); %Magnetic flux density and critical current
11 %% Geometry creation: lines of current are located at the points (Rx,Ry) %%
12 xRange=(1-m:2:m-1)*sw/2/m;               % Span of values for x coordinate
13 Rx=[repmat(xRange-(rg+sw)/2,[1 ns/2]),repmat(xRange+(rg+sw)/2,[1 ns/2])];
14 yRange=((2-ns):4:(ns-2))*sg/4;            % Span of values for y coordinate
15 Ry=[reshape(repmat(yRange,m,1),1,[]),reshape(repmat(yRange,m,1),1,[])];
16 %%       Definition of auxiliary variables for field calculation        %%
17 r2=bsxfun(@minus,Rx,Rx').^2+bsxfun(@minus,Ry,Ry').^2;% r2=(x-x')^2+(y-y')^2
18 xn=bsxfun(@minus,Rx,Rx')./r2; xn(isnan(xn))=0;     % if(r2>0,(x-x')/r2,0)
19 yn=bsxfun(@minus,Ry,Ry')./r2; yn(isnan(yn))=0;     % if(r2>0,(y-y')/r2,0)
20 %%                          Iterative solution                          %%
21 while (abs(max(P)-1)*s+abs(E/Ec-1)*(1-s)>tolp)        % Ic criteria loop
22 err = 1;                                              % Error reset
```

```matlab
23      while(err > tolIc)                                      % Self consistency loop
24          IcOld=Ic;
25          Ic=(sw*th/m)*(Jc0./(1+sqrt((k*Bx).^2+By.^2)./Bc).^b);  %local Ic(B)
26          P=reshape((I0./(reshape(Ic,m,ns)'*ones(m,m)))',1,m*ns);   %P values
27          It=P.*Ic;                                           % Local current in strand
28          Bx=-mu0/(2*pi)*It*yn;                               % Magnetic flux density
29          By=mu0/(2*pi)*It*xn;
30          err=norm(IcOld - Ic); % Error. It compares old and new Ic estimates
31      end
32      E=Ec*abs(sum(P(1:m:end).^n)/ns);    % Average electric field in cable
33      I0=2*I0/(s*(1+max(P))+(1+(E/Ec)^(1/n))*(1-s)); %Net current in strand
34  end
35  %%              Post Processing: Output data and plotting                %%
36  fprintf('Ic(cable)= %0.2f A (%s criteria)\n',ns*I0,c);
37  fprintf('Avg(E)= %0.6f microV/cm\n', E/1e-4);
38  fprintf('max(P)= %0.6f\n',max(P));
39  X=reshape(Rx,ns,[])/1e-3;
40  Y=reshape(Ry,ns,[])/1e-3;
41  Z=(ns/sw/th)*reshape(It,ns,[]);
42  mesh(X,Y,Z,'MeshStyle','column','FaceColor','none','LineWidth',3); % J(x,y)
43  xlabel('x [mm]'); ylabel('y[mm]'); zlabel('J [A/m^2]');
44  colorbar;
```

ACKNOWLEDGMENT

The authors are very thankful to Prof. Frédéric Hecht, developer of FreeFEM++ for his technical support in the implementation of the model presented here.

TABLE I
PARAMETERS OF THE ROEBEL CABLE CONSIDERED IN SECTION III A

| Parameter | Value | Description |
|---|---|---|
| $ns$ | 10 | Number of strands in Roebel cable |
| $sw$ | 1.8 mm | Strand's width |
| $th$ | 1 µm | Strand's thickness |
| $sg$ | 0.1 mm | Vertical distance between strands |
| $rg$ | 0.4 mm | Horizontal separation between stacks |
| $n$ | 21 | Exponent in the $E$-$J$ power-law relation |
| $m$ | 100 | No of current lines used to discretize each tape |

TABLE II
COMPARISON OF RESULTS GIVEN BY DIFFERENT IMPLEMENTATIONS

| Software | $Ic_{MAX}$ (A) | $Ic_{AVG}$ (A) | $ct_{MAX}$ (s) | $ct_{AVG}$ (s) |
|---|---|---|---|---|
| *Comsol* | 534.65 | 538.93 | 6.00 | 6.00 |
| *FreeFem++* | 535.76 | 537.12 | 34.30 | 45.01 |
| *Matlab* | 535.83 | 539.25 | 0.10 | 0.10 |
| *Octave* | 535.83 | 539.25 | 0.13 | 0.13 |

Ic=critical current, ct=computing time, the MAX and AVG labels indicate the criterion used. Computing times were measured excluding the part of the code that generates the plots.

TABLE III
COMPARISON OF $I_C$ OF A 12 MM-WIDE ROEBEL CABLE OBTAINED WITH DIFFERENT ANGULAR DEPENDENCES AND DIFFERENT CRITERIA.

| Software | $Ic_{MAX}$ (A) | $Ic_{AVG}$ (A) |
|---|---|---|
| *Precise $Jc(B,\theta)$* | 1005 | 1035 |
| *Simplified $Jc(|B|)$* | 1045 | 1067 |

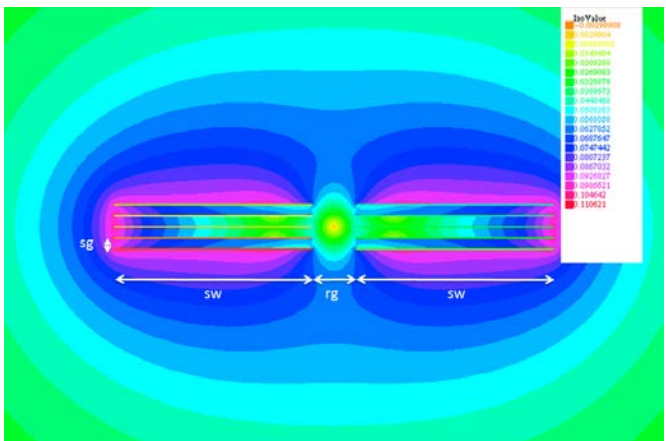

Fig. 1. Typical magnetic flux density distribution [T] in the Roebel cable considered in section III A.

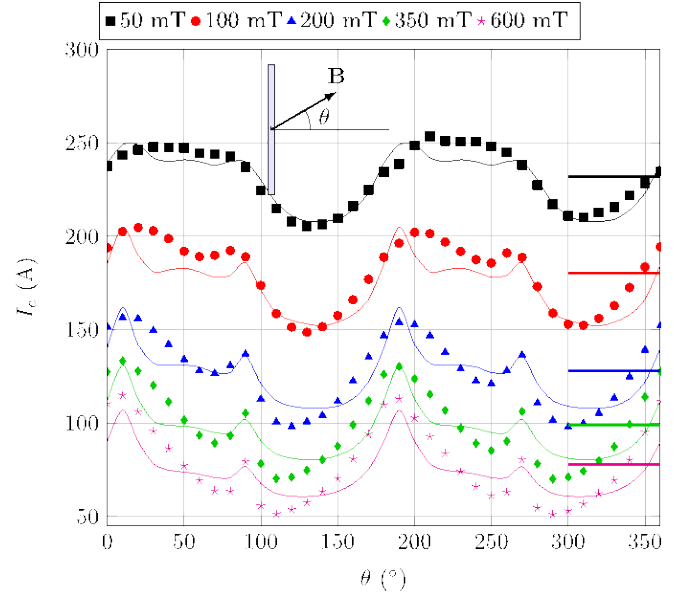

Figure 2. Angular dependence of $I_c$ for a 12 mm wide sample with artificial pinning. The horizontal lines represent the average value used for the model without angular dependence.

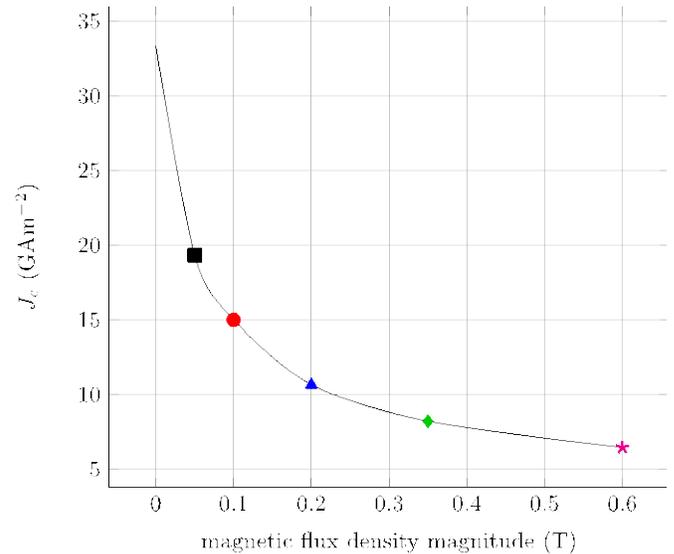

Figure 3. Simplified $J_c(|B|)$ dependence obtaining using the average values of Figure 2.